\documentclass[reprint, aps, prb, superscriptaddress]{revtex4-2}
\usepackage[version=3]{mhchem}
\usepackage{graphicx} 
\usepackage{float}
\usepackage{indentfirst}
\usepackage[hidelinks]{hyperref}
\usepackage{caption}
\usepackage{tablefootnote}
\DeclareCaptionLabelFormat{boldlabel}{\textbf{#1 #2}}
\usepackage{booktabs}
\usepackage{placeins}
\usepackage{multirow}
\usepackage{makecell}
\usepackage{tabularx}
\usepackage{adjustbox}
\usepackage[table]{xcolor}
\usepackage{hhline}
\usepackage{comment}

\definecolor{palegray}{gray}{0.95}
\definecolor{darkgray}{gray}{0.85}

\newcolumntype{g}{>{\columncolor{palegray}}c}
\newcolumntype{d}{>{\columncolor{darkgray}}c}

\begin{document}

\title{Adsorption energies are necessary but not sufficient to identify good catalysts}

\author{Shahana Chatterjee}
\affiliation{Mila, Montr\'eal, Canada}
\author{Alexander Davis}
\affiliation
{University of Toronto, Toronto, Canada}
\author{Lena Podina}
\affiliation{Mila, Montr\'eal, Canada}
\affiliation
{University of Waterloo, Waterloo, Canada}
\author{Divya Sharma}
\affiliation{Mila, Montr\'eal, Canada}
\affiliation
{Universt\'e de Montr\'eal, Montr\'eal, Canada}
\author{Yoshua Bengio}
\affiliation
{Mila, Montr\'eal, Canada}
\affiliation
{Universt\'e de Montr\'eal, Montr\'eal, Canada}
\author{Alexandre Duval}
\affiliation
{Entalpic, Paris, France}
\author{Oleksandr Voznyy}
\affiliation
{University of Toronto, Toronto, Canada}
\author{Alex Hern\'andez-Garcia}
\affiliation
{Mila, Montr\'eal, Canada}
\affiliation
{Universt\'e de Montr\'eal, Montr\'eal, Canada}
\author{David Rolnick}
\affiliation
{Mila, Montr\'eal, Canada}
\affiliation
{McGill University, Montr\'eal, Canada}
\author{F\'elix Therrien}
\affiliation
{Mila, Montr\'eal, Canada}
\email{felix.therrien@mila.quebec}

\keywords{Heterogeneous Catalysis, Electrocatalysis, Machine Learning, Adsorption Energy, Hydrogen Evolution Reaction, Oxygen Evolution Reaction, Open Catalyst Project, Thermodynamic Overpotential}

\begin{abstract}
 As a core technology for green chemical synthesis and electrochemical energy storage, electrocatalysis is central to decarbonization strategies aimed at combating climate change. In this context, computational and machine learning driven catalyst discovery has emerged as a major research focus. These approaches frequently use the thermodynamic overpotential, calculated from adsorption free energies of reaction intermediates, as a key parameter in their analysis. In this paper, we explore the large-scale applicability of such overpotential estimates for identifying good catalyst candidates by using datasets from the Open Catalyst Project (OC20 and OC22). We start by quantifying the uncertainty in predicting adsorption energies using \textit{ab initio} methods and find that $\sim$0.3-0.5 eV is a conservative estimate for a single adsorption energy prediction. We then compute the overpotential of all materials in the OC20 and OC22 datasets for the hydrogen and oxygen evolution reactions. We find that while the overpotential allows the identification of known good catalysts such as platinum and iridium oxides, the uncertainty is large enough to misclassify a broad fraction of the datasets as ``good'', which limits its value as a screening criterion. These results question the reliance on overpotential estimation as a primary evaluation metric to sort through catalyst candidates and calls for a shift in focus in the computational catalysis and machine learning communities towards other metrics such as synthesizability, stability, lifetime or affordability.
\end{abstract}

\maketitle

\section{Introduction}

The transition to a low-carbon economy hinges in part on advances in catalytic processes such as green hydrogen production and the conversion of atmospheric carbon dioxide into fuels to decarbonize sectors lacking sustainable alternatives \cite{masa_electrocatalysis_2020, sherrell_electrocatalysis_2024, oliveira2021green}. Consequently, the design or discovery of stable, efficient, sustainable and cost-effective catalysts is attracting growing interest from the computational materials science and machine learning communities \cite{norskov_fundamental_2014, mai_machine_2022, zitnick_introduction_2020}. 

Assessing and predicting what a good catalyst constitutes is central to that endeavor. The overpotential of an electrocatalytic reaction, which strongly correlates to catalytic activity, is widely used to  measure catalytic performance. In a method established by 
Nørskov and co-workers, the thermodynamic overpotential ($\eta$) can be computed as a function of the adsorption free energies of reaction intermediates, typically obtained with density functional theory (DFT) calculations \cite{norskov_origin_2004,  norskov_trends_2005, rossmeisl_electrolysis_2005, norskov_fundamental_2014}. This approach assumes that the activation energy of each reaction step correlates with the difference in adsorption energies between its reactants and products, as described by the Bronsted–Evans–Polanyi (BEP) relationship. Using this approach, adsorption energy calculations have been used both to successfully explain experimental observations and to predict trends in catalytic efficiency across small sets of selected materials \cite{norskov_trends_2005, rossmeisl_electrolysis_2005, koper_thermodynamic_2011, man_universality_2011, danilovic_enhancing_2012, divanis_oxygen_2020, zhang_adsorption_2022, broderick_identifying_2023, jones_toward_2024}.

Recently, the Open Catalyst Project has drawn significant attention to the field of catalyst materials discovery with the release of several large-scale datasets of adsorption energy calculations \cite{chanussot_open_2021, tran_open_2023, abed_open_2024}. Their stated goal is to accelerate the discovery of new catalyst materials through the use of machine learning (ML) \cite{zitnick_introduction_2020}. In practice, the rise of high-throughput ML-based catalyst screening has made Nørskov-style overpotential calculations the default way to assess catalytic performance.

ML models trained on these large datasets can not only speed up adsorption energy calculations, but can also open up the possibility of generating new materials. Several methods have emerged to generate realistic stable crystal structures with specific properties \cite{hernandez_crystal-gfn_2023, zeni2023mattergen, levysymmcd, merchant2023scaling}, including catalysts specifically \cite{podina2025catalyst}. In this inverse design paradigm, materials are generated and often optimized according to a certain useful property. When it comes to discovering new high performance catalysts, overpotential seems like a natural optimization objective, but can it really serve as a purely predictive, high-throughput metric for identifying the best candidates?

In this paper, we test whether overpotential is indeed a suitable metric to identify the best catalysts for these reactions. To do so, we use existing DFT data from the OC20\cite{chanussot_open_2021} and OC22\cite{tran_open_2023} datasets to compute and analyze the overpotential for two major electrocatalytic reactions: the hydrogen and oxygen evolution reactions (HER and OER). We start by quantifying the uncertainty of DFT-based adsorption free energy calculations and its effect on the predicted overpotential.
Then, we compute the overpotentials and filter all materials in both datasets according to this metric. We find that very large portions of the datasets are predicted as ideal catalysts within the uncertainty of the estimation. Although good catalysts such as platinum (for HER) can be correctly identified, overpotential lacks selectivity to pinpoint top performing materials. We discuss the implications of this finding and suggest that Pourbaix stability might be a much more selective filter.

\section{Results}

\subsection{Quantifying uncertainty in adsorption free energy calculations} \label{seq:uncertainty}

The first step in assessing the usefulness of overpotential as a screening metric is to quantify the uncertainty associated with its prediction. Analogous to resolution in optics (Rayleigh criterion), uncertainty determines the smallest difference in the predicted overpotential that can reliably distinguish the performance of two materials. In other words, if two catalyst surfaces have predicted overpotential that are within each other's uncertainty, it is not possible to confidently predict which one is better. A lower uncertainty therefore corresponds to a higher ability to discriminate small variations in performance. The closer the predicted overpotential of various materials is to each other, the smaller the uncertainty needs to be to meaningfully order them in terms of performance.




The adsorption free energies $\Delta G_{ads}$ of the HER and OER intermediates are calculated at standard pressure and temperature (STP, 298.15 K, 1 bar, pH = 0). This involves first computing the adsorption energy $\Delta E_{ads}$ at 0 K with DFT and then adding a Gibbs free energy correction term $\Delta G_{corr}$ to it.
\begin{equation} 
\Delta G_{ads} \text{(STP)} = \Delta E_{ads}\text{(0 K)} + \Delta G_{corr} 
\end{equation}

The uncertainty ($u$) on $\Delta G_{ads}$ is the sum of the uncertainties on $\Delta E_{ads}$ and $\Delta G_{corr}$ ($u_{ads}$ and $u_{corr}$ respectively). 


Calculating $\Delta E_{ads}$ requires making choices about DFT parameters such as k-point density, energy cutoff, and convergence criteria, as well as assumptions about the physical system, including surface terminations, coverages and adsorption sites. Literature practices vary widely, and while parameters can be tuned to reproduce experiments, predictive studies cannot rely on such adjustments.

To estimate the uncertainty of adsorption energy $\Delta E_{*H}$ in the context of HER, we evaluated $\Delta E_{*H}$ for nine metal surfaces while varying six different parameters within ranges of values found in the literature.  These sets of parameters all represent reasonable choices to represent the system without any prior knowledge. We then recorded the maximum difference in $\Delta E_{*H}$ that could be obtained across all sets of parameters and calculations from the literature (when available). We call this uncertainty $\mathbf{u_{ads}^\textbf{(param)}}$ and its value for HER ($\Delta E_{*H}$) is reported in Table~\ref{tab:uncertainty_summary}. See Section~\ref{sec:her_calcs} in the SI for a detailed analysis of these calculations.

For OER, we computed adsorption energies of *OH, *O, and *OOH on four oxide surfaces while varying five different parameters. Then, for each adsorbate combination we recorded $u_{ads}^\text{(param)}$. The range of values for the three OER adsorbates is reported in in Table~\ref{tab:uncertainty_summary}. See Section~\ref{sec:oer_calcs} in the SI for a detailed analysis of these calculations.


We also quantified statistically the effect of some of these parameters (surface termination, coverage, and adsorption site) by measuring the variance of calculations that used identical bulk structures and Miller indices in the OC20 and OC22 datasets. For example, if two calculations in OC22 have the same slab identifier and the same number of adsorbates we know that the difference in energy between these two calculations is likely due to the choice of the site. Here, we use the average root mean squared distance between calculated energies and the mean of each set of structures with corresponding parameters as a measure of uncertainty. We call it $\mathbf{u_{ads}^\textbf{(stats)}}$. Its values for HER on OC20 and OER on OC22 are reported in Table~\ref{tab:uncertainty_summary}. See Section~\ref{sec:stats} in the SI for a detailed a analysis of these calculations.


Finally, in addition to the uncertainty on $\Delta E_{ads}$, for each adsorbate (*H, *OH, *O, and *OOH), we quantified how much Gibbs free energy corrections vary across several materials. Gibbs free energy corrections are computationally expensive and are almost always assumed to be constant across surfaces and materials. Therefore, any variation across different materials of $\Delta G_{corr}$ adds to the uncertainty of the adsorption free energy. In this case, for each adsorbate, $\mathbf{u_{corr}}$ is the maximum difference between our calculated $\Delta G_{corr}$ and the constant $\Delta G_{corr}$ used in OC20 and OC22. The values for HER ($\Delta E_{*H}$) and a range for OER are reported in Table~\ref{tab:uncertainty_summary}. See Section~\ref{sec:gibbs_corr} in the SI for a detailed analysis of these calculations.


\begin{table}
    \centering
    \caption{Summary of adsorption free energy ($\Delta G_{ads}$) uncertainty quantification in the context of HER and OER. All values are in eV.}
    \label{tab:uncertainty_summary}
    \begin{tabular}{c@{\hspace{8pt}}ccccccc}
    \toprule
    Reaction & ($u_{ads}^\text{(param)}$ & or & $u_{ads}^\text{(stats)}$) & $+$ & $u_{corr}$ & $>$ & $u$ \\
    \midrule
    HER & 0.2 & & 0.3 & & 0.1 & & \textbf{0.3} \\
    OER & 0.5 & & 0.4-0.7 & & 0.2-0.3 & & \textbf{0.5} \\
    \bottomrule
    \end{tabular}
\end{table}


The quantities in Table~\ref{tab:uncertainty_summary} are estimates of different components of the uncertainty. $u_{ads}^\text{(param)}$ and $u_{ads}^\text{(stats)}$ represent two ways of estimating $u_{ads}$ that should be compounded with $u_{corr}$ to obtain a total uncertainty.
 
Given the values in Table~\ref{tab:uncertainty_summary}, we propose to use \textbf{0.3~eV} as a conservative (lower bound) estimate of the uncertainty of a hydrogen adsorption free energy calculation. Looking at values in Table~\ref{tab:uncertainty_summary} and given that bad terminations in OC22 might have an effect on $u_{ads}^\text{(stats)}$, we propose \textbf{0.5~eV} as a conservative estimate of the uncertainty of adsorption free energy calculations in the context of OER. These are the value we will use herein. 


\subsection{Dataset Analysis}

Now that we have established reasonable bounds for the adsorption free energy uncertainty we can evaluate if the overpotential can serve as a relevant metric to classify materials in the OC20 and OC22 datasets. To compute the thermodynamic overpotential $\eta$ from adsorption energies in acidic conditions we use the following formula:

\begin{equation}\label{eq:eta}
 \eta = \max_{0 < i \leq n} \{{\Delta G}_i - {\Delta G}_{i-1}\} - \frac{{\Delta G}_n}{n},     
\end{equation}

\noindent where $n$ is the total number of electrons in the reaction and $\{{\Delta G}_i\}$ are the adsorption free energies at each intermediate step $i$ with respect to the reactant (water in this case). For each reaction, ${\Delta G}_{0} = {\Delta G}_{\text{reactants}} \equiv 0$ and ${\Delta G}_n = {\Delta G}_{\text{products}} = {\Delta G}_\text{reaction}$. In the case of HER, we have
\begin{align*}
{\Delta G}_{0} &= {\Delta G}_{H_20} \equiv 0, \\
{\Delta G}_{1} &= {\Delta G}_{*H}, \\
{\Delta G}_{2} &= {\Delta G}_{H_2} \equiv 0.
\end{align*}

In this case, the free energy of the reaction is zero because of the computational hydrogen electrode (CHE) assumption that ${\Delta G}_{H_2} \equiv 2{\Delta G}_{*H}$. For HER, Equation~\ref{eq:eta} becomes simply equivalent to the absolute value of the ${\Delta G}_{*H}$. The OER reaction is slightly more complex with 4 intermediate steps: 

\begin{align*}
    {\Delta G}_{0} &= {\Delta G}_{H_20} \equiv 0, \\ 
    {\Delta G}_{1} &= {\Delta G}_{*OH}, \\ 
    {\Delta G}_{2} &= {\Delta G}_{*O}, \\ 
    {\Delta G}_{3} &= {\Delta G}_{*OOH}, \\
    {\Delta G}_{4} &= {\Delta G}_{O_2} \equiv 4.92.
\end{align*}

See the SI and methods for more details about the reaction steps and the details of these calculations.
 
\subsubsection{OC20 and the Hydrogen Evolution Reaction}

\begin{figure*}
    \centering
    \includegraphics[width=0.7\linewidth]{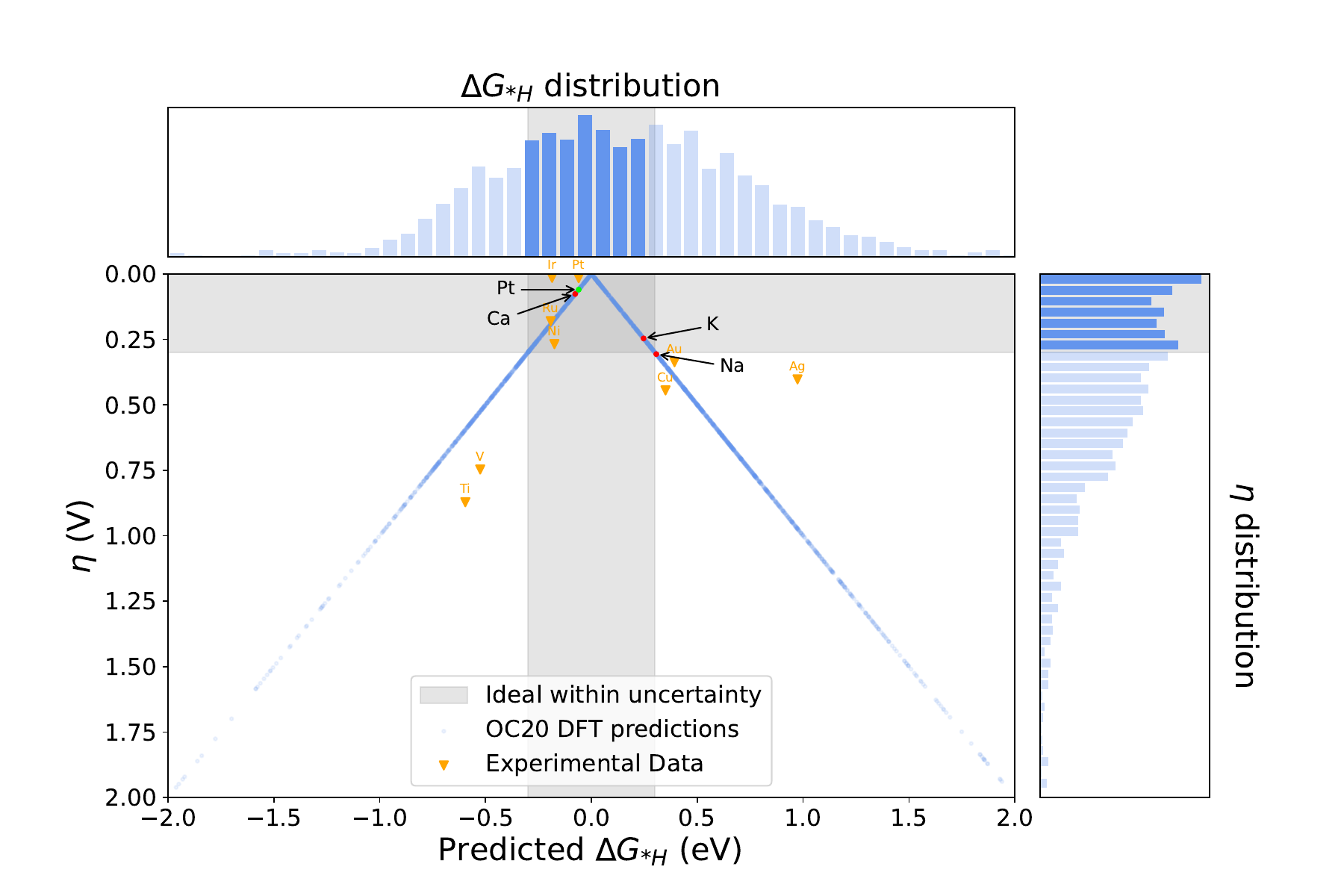}
    \caption{Overpotential as a function of calculated hydrogen adsorption free energy (${\Delta G}_{*H}$) in the OC20 dataset. Experimental data is from \citet{danilovic_enhancing_2012}}
    \label{fig:her}
\end{figure*}

We considered all surfaces in OC20 with *H adsorbates and filtered out surfaces marked as having anomalies, to obtain a total of 2402 surfaces covering 1937 different bulk crystals. We obtained their adsorption free energies by applying a constant Gibbs free energy correction of 0.24~eV from \citet{norskov_trends_2005} as is customary in the literature. Results are displayed in Figure~\ref{fig:her} in the form of a volcano plot between ${\Delta G}_{*H}$ and $\eta$ which, in this case, is simply its absolute value.

Ideal catalysts should have an overpotential as small as possible and should be at the upper center of this plot. The shaded area shows the region that is ideal within the uncertainty that we determined in Section~\ref{seq:uncertainty}: 0.3~eV. It is evident in Figure~\ref{fig:her} that the top of the ${\Delta G}_{*H}$ distribution is centered around 0 V and that a large portion of catalyst surfaces in OC20 are predicted to be ``perfect HER catalyst'' within uncertainty. In fact, 37\% of surfaces in OC20 fall within this uncertainty. Yet, materials in OC20 were chosen randomly from the Materials Project Database \cite{jain2013commentary} and there is no expectation that this dataset would be biased in any way towards good catalysts (see SI about exactly how materials were selected in these datasets). This is not to say that $\eta$ as calculated in OC20 does not correlate to reality. For example, orange points in Figure~\ref{fig:her} show at least some correlation between experimentally measured overpotential from \citet{danilovic_enhancing_2012} and OC20 calculations. However, it is unlikely that the Materials Project and OC20 contain hundreds of undiscovered catalysts with similar performance as Pt. Looking at some of the materials that fall within the uncertainty, one can find materials such as Na, K or Ca that are highly reactive and would quickly disintegrate or explode in solution. Although it might be true that a hypothetically stable Ca surface might be an excellent HER catalyst, its lack of stability in water ensures that it cannot be used in practice. Overpotential may be necessary to identify a good catalyst, but because it is so easy to find catalysts with ideal $\eta$ in the Materials Project it is not sufficient to pin-point which catalyst is actually good.

\subsubsection{OC22 and the oxygen evolution reaction}

\begin{figure*}
    \centering
    \includegraphics[width=0.7\linewidth]{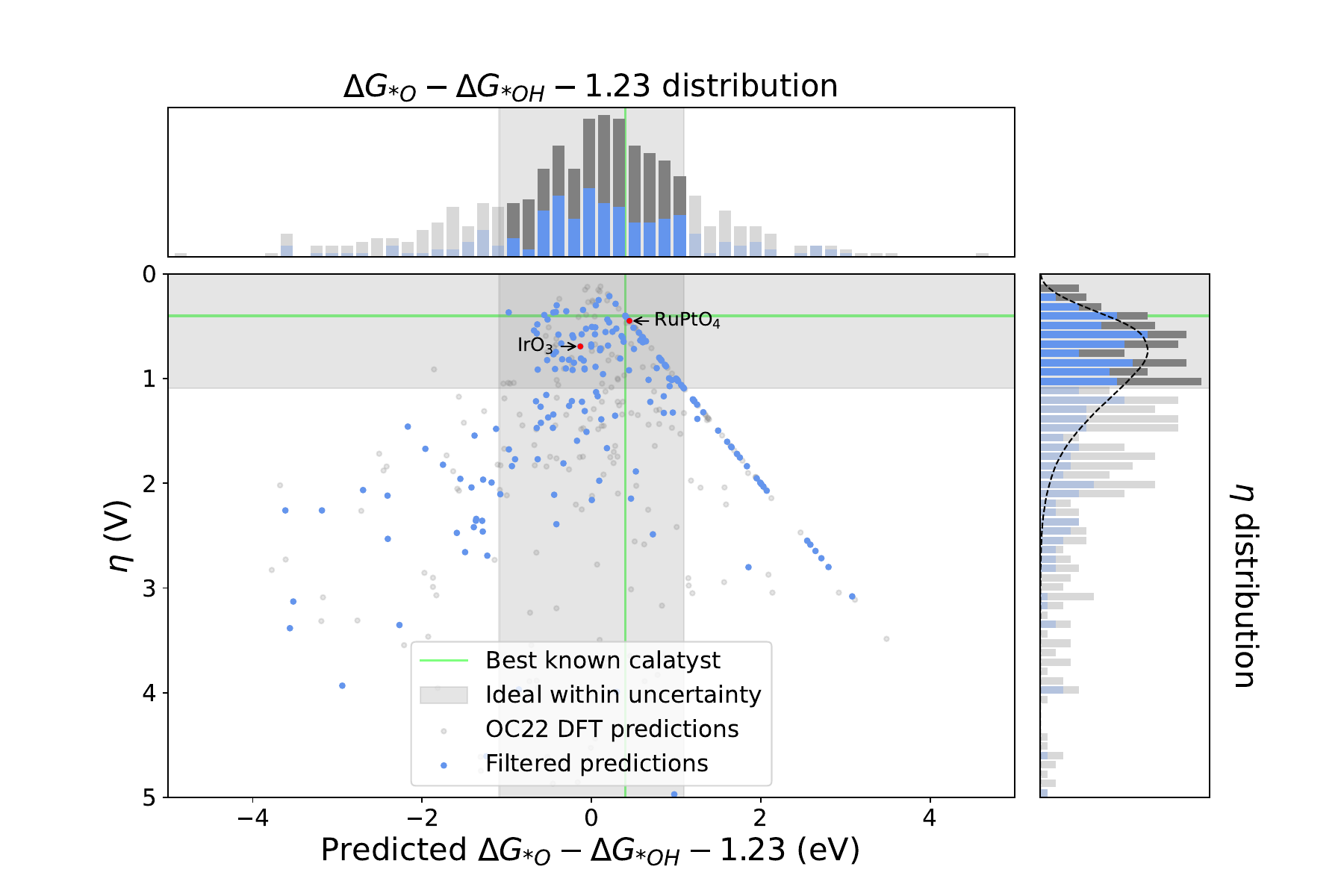}
    \caption{Overpotential as a function of the typical rate limiting step (${\Delta G}_{*O}$ - ${\Delta G}_{*OH}$) in the OC22 dataset. The dotted line is the distribution of calculated $\eta$'s for the best known catalyst under the assumption that adsorption free energy calculations are normally distributed with a standard deviation of 0.5~eV.}
    \label{fig:oer}
\end{figure*}

Similarly, we studied all surfaces in OC22 where energies were calculated for *OH, *O and *OOH. There were a total of 491 surfaces that met this criterion corresponding to 359 materials. Upon analyzing the data, we realized that some calculations involved unstable terminations for reaction intermediates *OH and *OOH that caused them to dissociate. We went through all surfaces and measured distances and angles between atoms of the *OH and *OOH adsorbates after relaxation, discarding all calculations that did not meet our criteria. For *OH, we required the O–H bond length to fall between 0.9 and 1.1~\AA{}. For *OOH, we applied the same 0.9–1.1~\AA{} threshold to the O–H bond of the terminal hydroxyl group, and required the O–O distance of the O–O–H unit to lie between 1.3 and 1.5~\AA{}, with an O–O–H angle between 95 and 115 degrees.  After this filtering, there were 187 surfaces from 147 materials remaining. We applied fixed, adsorbate-specific Gibbs free energy corrections, one constant each for *O, *OH, and *OOH, as calculated in OC22 \cite{tran_open_2023}. We computed the overpotential $\eta$ using Equation~\ref{eq:eta} for all surfaces; results for both filtered and unfiltered data are presented in Figure~\ref{fig:oer}.

In Figure~\ref{fig:oer}, we plot the OER overpotential with respect to the reaction free energy for the *OH to *O step. This step is considered important because of the existence of a scaling relationship between ${\Delta G}_{*OH}$ and ${\Delta G}_{*O}$ \cite{man_universality_2011} that correlates relatively strongly with $\eta$. The difference in  ${\Delta G}_{*O}$ and ${\Delta G}_{*OH}$ values equals the reaction free energy for this step and, because of this scaling relationship, should be around 1.6 eV for an ideal catalyst. Hence, ideal catalysts are expected to be at the top (low $\eta$) and at the center. Because it is a difference between two free energies, the total uncertainty is simply the sum of the uncertainties: 1~eV. It is once again evident by looking at the top of Figure~\ref{fig:oer} that the distribution is centered around the ideal value and that a large portion of the dataset is within the ideal zone.

Propagating the uncertainty of the adsorption energy to $\eta$ is not as straightforward for the OER. We know that the best known catalysts IrO$_2$ and RuO$_2$ have overpotentials of about 0.4~eV because they are limited by the scaling relationship. So we know that their adsorption energies are approximately the following: ${\Delta G}_{*OH}$ = 1.23~eV, ${\Delta G}_{*O}$ = 2.83~eV and ${\Delta G}_{*OOH}$ = 4.43~eV. Using these values in Equation~\ref{eq:eta} yields $\eta_{ideal}$ = 0.37~eV. Now, assuming that these are the true values and treating the uncertainty from Section~\ref{seq:uncertainty} as statistical uncertainty i.e., normal distributions centered on each ${\Delta G}$ with a standard deviation of 0.5~eV we can obtain the resulting distribution of $\eta$. This distribution is plotted in the right panel of Figure~\ref{fig:oer}. It is, in some way, the distribution of $\eta$ we expect if we were to compute $\eta$ for the ideal catalyst with multiple sets of parameters as in Table~\ref{tab:oer_experiments}. For normally distributed data, the uncertainty is often chosen to be one standard deviation; about 68\% of the data falls within that range. In our ideal distribution, the same proportion of the overpotentials (68\%) are within 1.09~eV. This is the threshold we will use to determine if a surface is as good as the best known catalyst ``within uncertainty''. In fact, applying this exact same procedure for the HER case where the distribution is normal would yield a threshold of 0.3~eV.

Using this threshold, we find that 34\% of all structures and 47\% of filtered structures are within uncertainty. Yet, once again, materials from OC22 are oxides chosen randomly from the Materials Project where there is no expectation of any bias towards good OER catalysts. The best known catalyst IrO$_2$ and RuO$_2$ were not randomly selected to be part of OC22, but similar materials such as IrO$_3$ and RuPtO$_4$ are predicted to have good performance suggesting once again that overpotential can correctly predict good catalysts. However, it is unlikely that almost half of the randomly selected materials in the OC22 dataset happen to be excellent catalysts. This suggests that overpotential is a weak metric to identify good catalysts. In other words, although it may or may not have good specificity (i.e. the ratio of correctly detected negatives to all actual negatives), it is definitely not very selective (i.e. it leads to many false positives).

Here, we did not consider the choice of the surface orientation as a factor in the prediction of $\eta$: all surfaces are represented separately in Figures\ref{fig:her} and \ref{fig:oer}. In practice, the goal is to obtain a prediction for the material as a whole. One could assume that the overpotential of a material is simply that of the best performing surface or they could use a more elaborate aggregation method based on methods such as the Wulff construction \cite{sanspeur2023wherewulff}. This could have an impact on the percentage of materials that are predicted as high performing in OC20 and OC22, but, once again, since Miller indices are chosen randomly, there is no reason to expect that this would shift the distribution one way or the other. 

\section{Discussion}

The analysis presented in the previous section does not indicate that adsorption energy based overpotential predictions are necessarily wrong. At best (scenario 1), the overpotential predictions are indeed correct and closely match the true values with many materials truly having low overpotentials. However, the fact that identifying cheaper and more efficient catalysts for water splitting remains an active area of research suggests that the hundreds of low $\eta$ materials in OC20 and OC22 are most likely not good catalysts for other reasons. In experimental studies, there is an implicit bias towards materials that \textit{can} work as catalysts. This is why no experimental potential data exist for Na, K and Ca; these metals react vigorously with water, forming hydrogen gas and hydroxides almost instantaneously. This could explain why there is a relatively good correlation between predicted and experimental overpotential in Figure~\ref{fig:her} for the literature \cite{norskov_trends_2005, man_universality_2011, ostergaard2022predicting} dataset. In this context, a low overpotential appears to be a necessary condition for catalytic activity, but because it is relatively easy to satisfy, it is not sufficient for identifying truly effective catalysts. This idea has already been discussed in the context of HER by \citet{broderick_identifying_2023} stating: ``we discover that favorable adsorption energies are a necessary condition for experimental activity, but other factors often determine trends in practice.''
\begin{figure*}
    \centering
    \includegraphics[width=0.9\linewidth]{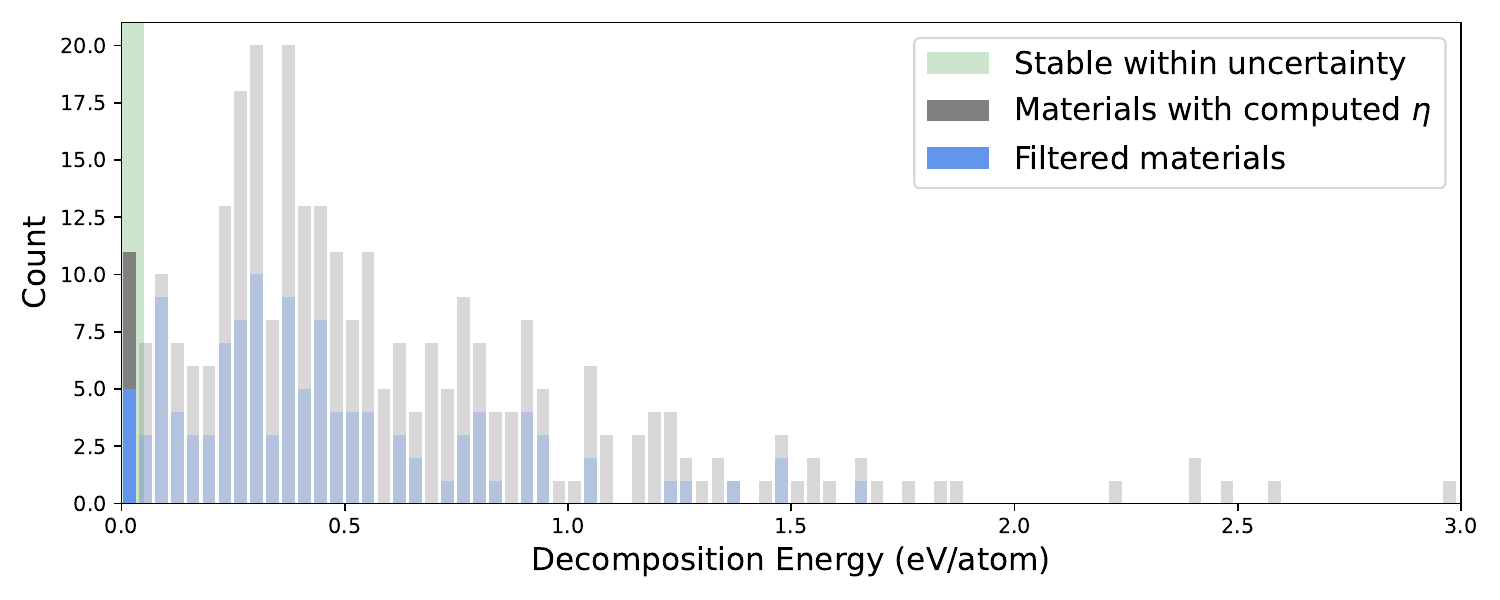}
    \caption{Decomposition energy for materials in OC22 that have at least one surface with all three OER adsorption energies ($\Delta G_{*OH}$, $\Delta G_{*O}$, $\Delta G_{*OOH}$) and are part of the Materials Project database}
    \label{fig:stability}
\end{figure*}

At worst (scenario 2), adsorption energy-based overpotential predictions in OC20 and OC22 are inaccurate for a large portion of the materials in these datasets. This could be because of an inaccurate representation of coverage, adsorption site, termination, surface orientation, etc. and how different combinations of these physical parameters are aggregated into a single prediction. In this scenario there might be a way to obtain more accurate predictions with a better description of the system. However, \citet{abed_open_2024} found that when linearly fitting experimental potentials for HER with computational data, descriptive features of the materials alone fit the experimental data almost as well as DFT-predicted adsorption energies across different aggregation strategies suggesting a relatively weak correlation between adsorption energy and experiment regardless of how sites and surfaces are treated. 

While the predictions in OC20 and OC22 rely on the BEP principle, it is possible that computing actual transition states and obtaining activation energies would shift $\eta$ distributions down without affecting actual good catalysts (e.g., Pt). In other words, it is possible that good catalysts stand out because of their activation energies, not their adsorption (reaction) energies. Computing transition states in a high-throughput context is computationally expensive, but ML-based methods such as CatTSunami\cite{wander2025cattsunami} can make this process significantly more affordable. It is to be determined if systematically calculating such activation energies and taking into account their uncertainty on the entire dataset would make predicted $\eta$ more selective.

In both scenarios, adsorption energy based calculations alone are weak filters for what constitutes a good catalyst. There are several examples of computational and machine learning studies in the literature that focus on other metrics first. For example, \cite{abed2024pourbaix} screened through thousands of materials using a machine learning model of Pourbaix stability or Pourbaix decomposition energy: the energy of a material above the most stable phase at specific pH and potential. \cite{tran2024rational} also used Pourbaix stability as a first filter, preceding overpotential, to identify good catalysts in OC22.

We computed the Pourbaix decomposition energy at a pH of 0 (acidic conditions) and a potential of 1.63~V for all 305 materials for which we had obtained $\eta$ in Figure~\ref{fig:oer} that were also part of the Materials project. The distribution of their decomposition energy is shown in Figure~\ref{fig:stability}. We use an uncertainty of 50~meV, because energy differences for bulk oxides in the Materials Project are estimated to have a standard deviation as low as 24~meV \cite{hautier2012accuracy}. Within that uncertainty, less than 5\% of bulk structures are predicted to be stable. Even using a much more permissive 200~meV cutoff, it is only 14\% of structures that are predicted stable. Decomposition energy being a bulk property that, in many cases, only requires a single energy calculation, it may constitute a more relevant objective function for catalyst discovery than overpotential, especially in the context of generative crystal models. Hence, compared to adsorption energies, Pourbaix stability appears to be a better criteria for catalyst screening: using the adsorption energy metric $\eta$ yields a broad range of candidates while Pourbaix stability narrows the selection.

The Open Catalyst 2020 and 2022 datasets have become ubiquitous for computational catalysis research. Our paper shows how adsorption energies from these datasets and therefore models trained on this data may not be \textit{directly} useful to identify or generate good catalysts. However, we want to emphasize that (1) our analysis and our conclusions would obviously not have been possible without them, (2) they remain extremely valuable to train machine learning force fields and (3) they have brought remarkable attention to this field from the ML community. We hope that our paper can steer this community including the Open Catalyst Project towards more effective objectives for catalyst discovery.

\section{Conclusion}

Materials discovery requires well-defined objective functions, which, for high throughput screening must be theoretical in nature. For electro-catalytic reactions, the objective is often related to minimizing the predicated overpotential based on adsorption free energy calculations. Focusing on the hydrogen and oxygen evolution reactions and using DFT-calculated adsorption energy data from the Open Catalyst Project, we tested the viability of this approach.  

We first quantified the inherent uncertainty in DFT-calculated adsorption free energies by running various adsorption energy calculations with several sets of parameters and with statistical analyses of the OC20 and OC22 datasets. We found that 0.3~eV is a conservative estimate of the uncertainty of adsorption free energies on metal surfaces used for HER and that 0.5~eV is a conservative estimate of the uncertainty of adsorption free energies on oxide surfaces used for OER.

We computed the overpotential for the hydrogen and oxygen evolution reaction of all entries that had sufficient information in the Open Catalyst 2020 and Open Catalyst 2022 datasets respectively. We found that, according to the overpotential criterion, 37\% of catalyst surfaces in OC20 and 47\% of catalyst surfaces in OC22 would be considered to be as good or better than the best known catalysts for their respective reactions within our estimate of the uncertainty. Since the structures in these datasets are chosen randomly within the Materials Project database, there is no expected bias towards good catalysts. Therefore, we conclude that adsorption energy based overpotential cannot be used to selectively identify good catalysts.

These findings highlight the limitations of relying solely on the adsorption energy based thermodynamic overpotential and other similar metrics for catalyst screening. Factors such as catalyst stability, cost, synthesis feasibility, and environmental impact may represent stronger and cheaper filters that should be considered first. As a possible alternative, we find that Pourbaix stability is much more selective when applied to OC22, with less than 5\% of structures predicted to be stable. Using more effective objectives like stability can help prioritize better candidates and lead to more practical and successful catalyst discovery.

\section{Computational Methods}
All DFT calculations were performed with the VASP package (version 6.3.0) and $\Delta G_{corr}$ values were calculated using the VaspGibbs package \cite{therrien_vaspgibbs_2023}.

For the HER, DFT calculations were performed using five-layer slabs, with the top two layers relaxed and a 60 Å vacuum separating periodic images. Hydrogen adsorption was modeled at octahedral or octahedral-like sites on fcc$\{111\}$ and bcc$\{110\}$ surfaces, at coverages of 0.25 and 1 monolayer. Both RPBE and PBE exchange–correlation functionals were used with a plane-wave cutoff of 350 eV and a 4×4×1 k-point mesh. 

For the OER, slabs of perovskites SrAO$_3$$\{100\}$ (B = Ti or Co) were terminated with the B atoms, and *O, *OH, and *OOH intermediates were adsorbed directly atop these B sites. The slab contained three atomic layers, with only the top layer relaxed. For rutile oxides, MO$_2$$\{110\}$ (M = Ru or Ir) slabs were constructed with surface M atoms exposed, and adsorbates were placed on top of the M sites; in this case, the top layer of a four-layer slab was relaxed.  All calculations employed a plane-wave cutoff energy of 500 eV, spin polarization, and the RPBE or the PBE exchange–correlation functionals. 

For both the HER and the OER, dipole corrections were applied, and structural relaxations were carried out using the conjugate-gradient algorithm.

Pourbaix decomposition energies were computed using pymatgen's PourbaixDiagram feature.For each material, energies of materials and decomposition products required for the Pourbaix diagram were retrieved from the Materials Project as PourbaixEntry objects.
Then, pymatgen was used to create a room-temperature Pourbaix diagram as a PourbaixDiagram object, using the \texttt{filter\_solids} flag to remove unstable phases, and using the default ion concentration of $10^{-6}$ M. When these unstable phases are present in OC22, they can lead to negative Pourbaix energies. Therefore, we excluded structures that lead to negative energies from Figure~\ref{fig:stability}.  
The Pourbaix diagram was then sampled at a pH of 0 and a potential of 1.63 V to obtain the predicted decomposition energy.

\section{Code Availability}
The code to reproduce all figures and tables is available on our public repository \footnote{\href{https://github.com/chemicallyGeeky/OCdataset_analysis_for_HER_OER}{github.com/chemicallyGeeky/OCdataset\_analysis\_for\_HER\_OER}}.

\section{Acknowledgements}
S.C. gratefully acknowledges the late Prof. Meenakshi Chatterjee for the valuable discussions in 2022 that guided her toward exploring machine learning, an area she might not have otherwise pursued. A.D. and O.V. acknowledge support from the Alliance for AI-Accelerated Materials Discovery (A3MD). Authors affiliated with Mila, ackowledge the support of Samsung.

\section{Supporting Information}
Details on adsorption energy calculations with DFT, derivation of objective functions for the HER and the OER from respective reaction mechanisms, supporting figures and tables.

\clearpage
\bibliography{references}

\onecolumngrid
\renewcommand{\thetable}{S\arabic{table}}
\setcounter{table}{0}
\renewcommand{\thefigure}{S\arabic{figure}}
\setcounter{figure}{0}
\renewcommand{\theequation}{S\arabic{equation}}
\setcounter{equation}{0}
\renewcommand{\thesection}{S\arabic{section}}
\setcounter{section}{0}

\clearpage

\begin{center}
\Large Supplemental Information
\end{center}
\bigskip
\suppressfloats[t]

\section {Methods and background}
\subsection{Adsorption energy calculations}

\begin{table}
\centering
    \caption{List of acronyms and symbols}
    \label{tab:symbols}
    \begin{tabular}{cl} 
    \toprule
        * & Empty surface site \\ 
        ads & Free adsorbate \\ 
        ads & Adsorbate on a surface site \\ 
        a$_{H^+}$ &  Activity of H$^+$ at a given pH \\ 
        C$_\text{p}$ &  Heat capacity at constant pressure \\ 
        $\eta$ &  Thermodynamic overpotential \\ 
        eV  &  Electron volt \\ 
        $G$  &  Gibbs' free energy  \\ 
        k$_\text{B}$ &  Boltzmann constant \\  
        \textit{P} &  Pressure \\ 
        \textit{S} &  Entropy \\ 
        STP &  Standard temperature and pressure (1 bar, 298.15 K) \\ 
         \textit{T} &  Temperature \\ 
         ZPE & Zero-point energy \\ 
         \bottomrule
    \end{tabular}
\end{table}

\begin{table}
    \centering
    \caption{Terminology used in adsorption energy calculations.}
    \label{tab:dftCalc}
    \begin{tabular}{cp{8cm}} 
    \toprule
        Slab &  A three-dimensional crystallographic model of an infinite  two-dimensional surface in adsorption energy calculations. \\ \midrule 
         Slab with adsorbate &  A slab with adsorbates on its surface. \\ \midrule 
         \textit{E} &  Refers to energy obtained from a DFT calculation.  This is the electronic energy of the system at 0 K, without any zero-point correction. \\ \midrule 
         $\Delta E_{ads}$ &  This is the adsorption energy of a slab with adsorbate, obtained from the \textit{E} values of the slab, slab with adsorbate and free adsorbate. It refers to the electronic energy contribution at 0 K (to the total adsorption energy). \\ \midrule 
         $\Delta G_{ads}$ &  This is the adsorption free energy of a slab with adsorbate at \textit{T}, \textit{P}.  \\ \midrule
         $\Delta G_{corr}$ &  The energy correction that must be added to $\Delta E_{ads}$
         to obtain $\Delta G_{ads}$. \\ 
         \bottomrule
    \end{tabular}
\end{table}

All adsorption and reaction energies are calculated at STP. Also, these values are in eV per species.  The following relationships have been used to calculate the adsorption energy at STP:

\begin{equation} \label{AE-si1} \Delta E_{ads} = E\text{(slab with adsorbate)} - E\text{(slab)} -E\text{(adsorbate)} \end{equation}
\begin{equation} \label{AE-si2} \Delta G_{ads} = \Delta E_{ads} + \Delta G_{corr} \end{equation}
where,
\begin{equation} \label{AE-si3} \Delta G_{corr} = G_{corr}\text{(slab with adsorbate)} -G_{corr}\text{(slab)} -G_{corr}\text{(adsorbate)} \end{equation}
with,
\begin{equation} \label{AE-si4}
    G_{corr} = ZPE + \int_{0}^{298.15}C_p(T, 1  \text{ bar})dT - 298.15*S(\text{STP})
\end{equation}

\subsubsection{Details of DFT calculations}  

All DFT calculations were performed with the VASP package (version 6.3.0) and $\Delta G_{corr}$ values were calculated using the VaspGibbs package \cite{therrien_vaspgibbs_2023}.  Final adsorption energy values were obtained using equations \ref{AE-si1} - \ref{AE-si4}.
 
\begin{table}
    \centering
    \caption{Default DFT settings for HER calculations}
    \label{tab:dftHER_table}
    \begin{tabular}{lp{4 cm}p{4cm}p{4cm}} 
    \toprule
         & Literature & OC20 & Our calculations \\
         \midrule
         surface orientation & $\{111\}$ (fcc) or $\{110\}$ (bcc) & varies & $\{111\}$ (fcc) or $\{110\}$ (bcc) \\ 
         slab construction & 5, top 2 relaxed & varies, for Pt$\{111\}$ 1 of 6 layers relaxed & 5, top 2 relaxed  \\ 
         adsorption site & $\{111\}$: octahedral, $\{110\}$: octahedral-like & Pt$\{111\}$: tetrahedral & $\{111\}$: octahedral, $\{110\}$: octahedral-like \\ 
         coverage & 0.25 or 1 monolayer & varies, 1/9 for Pt$\{111\}$ & 0.25 or 1 monolayer \\ 
         functional & RPBE & RPBE & RPBE or PBE \\ 
         ENCUT in eV &  350 & 350 & 350 \\ 
         KPOINTS & 4x4x1 & - & 4x4x1 \\ 
         dipole correction for slabs& on & off & on\\ 
         EDIFF & - & 1e$^{-4}$ &  1e$^{-5}$ \\ 
         ionic relaxation method & - & conjugate gradient & conjugate gradient\\ 
         EDIFFG & - & -0.03 eV/\AA & -0.01 eV/\AA \\ 
           NSW & - & 200 & 200 \\ 
    \bottomrule
    \end{tabular}
\end{table}

\begin{figure}
    \centering
    \includegraphics[width=0.75\linewidth]{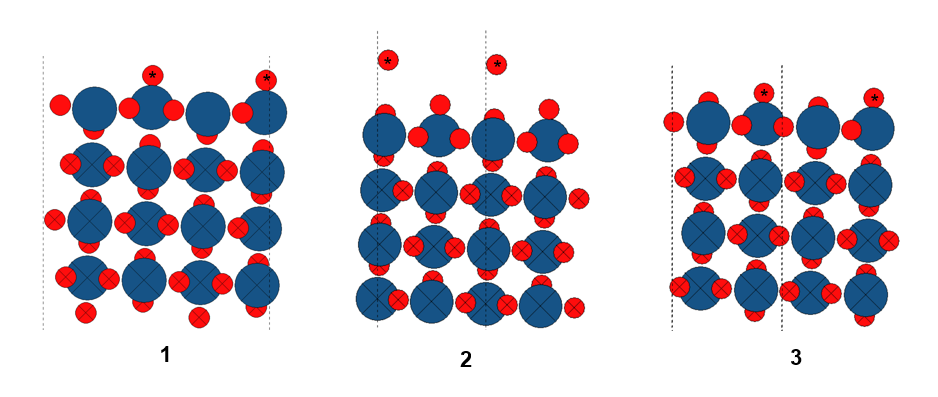}
    \caption{Side view of the IrO$_2$$\{$110$\}$ surface showing the surface terminations, the axis perpendicular to the surface points up. Blue: Ir, red: O; the starred O are the adsorbed O. }
    \label{fig:terminations}
\end{figure}

Adsorption energies for *O, *OH and *OOH were calculated for the $\{$100$\}$ surface for perovskite structures and the $\{$110$\}$ surface for rutile structures, and, different adsorption sites were explored with the RPBE functional.  For our calculations, the SrTiO$_3$ $\{$100$\}$ was cut such that Ti atoms were on top, the adsorbates were placed on top of these Ti atoms and the top layer was relaxed in the three-layered slab.  For IrO$_2$$\{$110$\}$, the adsorbates were placed on top of Ir, and the top layer was relaxed in the four-layered slab (Figure \ref{fig:terminations}).

\begin{table}
    \centering
    \caption{Default DFT settings for OER calculations}
    \label{tab:dftOER_table}
    \begin{tabular}{llll}
    \toprule
         & Literature & OC22 & Our calculations \\
         \midrule
         coverage & variable & variable & high \\ 
         ENCUT in eV &  500 & 500 & 500 \\ 
         KPOINTS & - & (30/a)x(30/b)x1 (density)& 4x4x1 \\ 
         dipole correction for slabs& - & - & on\\ 
         spin & on & on & on\\ 
         functional & RPBE & PBE & RPBE or PBE \\ 
         EDIFF & - & 1e$^{-4}$ &  1e$^{-5}$ \\ 
         ALGO & - & Fast &  Normal \\ 
         ionic relaxation method & conjugate gradient & conjugate gradient & conjugate gradient\\ 
          EDIFFG & -0.05 eV/\AA & -0.05 eV/\AA & -0.01 eV/\AA \\ 
           NSW & - & - & 200 \\ 
           \bottomrule
    \end{tabular}
\end{table}

\subsubsection{Adsorption energy calculations for the OC20 and OC22 datasets}
The energies provided in the OC20 dataset are $\Delta E_{ads}$ values (as `y-relaxed').  These were converted to $\Delta G_{ads}$ values according to equation \ref{AE-si2} using a $\Delta G_{corr}$ value of 0.24 eV \cite{norskov_trends_2005}.  Because some DFT settings were unavailable in the OC20 paper, we extracted and recalculated select structures using our own settings to verify our computational setup (Table \ref{tab:herDFTextracted}).

 Adsorption energy calculation from the OC22 Dataset for the OER: The energies provided in the OC20 dataset are the \textit{E} values of clean slabs, slabs with adsorbates and free adsorbates. Hence, $\Delta E_{ads}$ values were first calculated according equation \ref{AE-si1} followed by the calculation of $\Delta G_{ads}$ values according to equation \ref{AE-si2} with $\Delta G_{corr}$ values from the OC22 paper. \cite{tran_open_2023} 

 \begin{table}[H]
    \centering
    \caption{$\Delta E_{*H}$ value comparison between our calculations and OC20, for structures extracted from the OC20 dataset.}
    \label{tab:herDFTextracted}
    \begin{tabular}{lllll}
    \toprule
         & Materials project id & Composition & OC20 & Our calculations \\ 
         \midrule
         Pt$\{111\}$ & mp-126 & Pt54H1 & -0.30 & -0.30 \\ 
         Ca$\{111\}$ & mp-21 & Ca16H1 & -0.32 & -0.30  \\ 
         Nb$\{111\}$ & mp-75 & Nb54H1 & -0.31 & -0.31 \\ 
         \bottomrule    
    \end{tabular}
\end{table}

\subsection{Using adsorption energies for predicting catalytic efficiency}

 \subsubsection{BEP Relationship}  
 The Bronsted-Evans-Polanyi relationship, links the activation energy E$_a$ to the reaction energy $\Delta G$ as:
\begin{equation}
    E_a = \alpha \Delta G + \beta
\end{equation}
where $\alpha$ and $\beta$ are parameters.
For calculating the overpotentials of the HER and the OER, $\alpha$ and $\beta$ have implicit values of 1 and 0 respectively. 

\subsubsection{The thermodynamic overpotential for the HER}
 
 The HER reaction is a two-step reaction with an equilibrium potential of 0 V and a Volmer step followed by a Tafel or Heyrovsky step.

\begin{equation} 
\begin{aligned}
\text{Volmer step:} \qquad H^+ + e^- + * \rightarrow H*  \\
\text{Tafel step:} \qquad H* + H* \rightarrow H_2 + 2* \\
\text{Heyrovsky step:} \qquad H* + H^+ + e^- \rightarrow H_2 + * \\
\text{Overall:} \qquad 2 H^+ + 2 e^- \rightarrow H_2  
\end{aligned}
\end{equation}

Typically, the Volmer step is the potential determining step.  From the BEP principle, this potential determining step is assumed to be equivalent to the rate determining step. 

\subsubsection{The thermodynamic overpotential for the OER}
 
Acidic OER is a four step reaction.
\begin{equation}
\begin{aligned}
 \text{Step 1:} \qquad H_2O \text{ (liquid)} + * \xrightarrow {} HO* + H^+ + e^- \\
  \text{Step 2:} \qquad *OH \xrightarrow{} *O + H^+ + e^- \\
 \text{Step 3:} \qquad *O + H_2O \text{ (liquid)} \xrightarrow{} *OOH + H^+ + e^- \\
\text{Step 4:} \qquad*OOH \xrightarrow{} * + O_2 \text {(gas)} + H^+ + e^- \\
\text{Overall:} \qquad 2H_2O \text{(liquid)} \quad\xrightarrow[]{} \quad O_2\text{(gas)} + 4H^+ + 4e^-
\end{aligned}
\end{equation}
Typically step 2 or 3 is the potential determining step, which is also the rate determining step (from the BEP principle). 
The adsorption energies of the OER intermediates are defined as:

 \begin{equation}
 \begin{aligned}
 \Delta G_{*OH} &= G_{*OH} - G_* - (G_{H_2O} - \frac{1}{2}G_{H_2}) \\
 \Delta G_{*O} &= G_{*O} - G_* - (G_{H_2O} - G_{H_2})    \\
 \Delta G_{*OOH} &= G_{*OOH} - G_* - (2G_{H_2O} - \frac{3}{2} G_{H_2}) 
\end{aligned}
\end{equation}

Relationship between adsorption energies of *O, *OH and *OOH: As adsorbates *O, *OH and *OOH all bind to the catalyst surface via the $\text{O}$ atom, there exists a relationship between them. \cite{koper_thermodynamic_2011} 
\begin{equation} 
\begin{aligned}
\Delta G_{*OH} = 0.50* \Delta G_{*O} + 0.04 \\
 \Delta G_{*OOH} = 0.53*\Delta G_{*O} + 3.33 
 \end{aligned}
\end{equation}
Hence,
\begin{equation}    \Delta G_{*OOH} - \Delta G_{*OH} = 3.2 \ \text{eV} \end{equation}
For an ideal catalyst:
\begin{equation}    \Delta G_{*O} - \Delta G_{*OH} = 1.6 \ \text{eV} \end{equation}

\section{Quantifying uncertainty in the adsorption free energy calculations}
\subsection{HER calculations} \label{sec:her_calcs}

The hydrogen evolution reaction (HER) is a one-step reaction that requires only the adsorption energy of hydrogen ($\Delta E_{*H}$) for calculating the reaction overpotential. There are many parameters that need to be chosen when calculating $\Delta E_{*H}$, some lead to more or less accuracy in the DFT itself (k-point density, energy cutoff, convergence criteria, etc.) while others depend on one's knowledge or intuition of the actual physical system (coverage, termination, site, etc.). The choice of this set of parameters varies in the literature and is not always fully reported. When the goal is to explain or rationalize experimental results, the parameters can be adjusted to reflect experimental observations. For instance, the surface coverage may be tuned in the calculation so that the resulting overpotential matches the experimentally observed value. However, in a purely predictive context these types of adjustments are not possible and there does not exist a universal set of ``correct'' parameters. Moreover, when going through thousands of calculations, certain parameters are chosen considering resource usage. 

To quantify the effects of such computational parameters, we calculated hydrogen adsorption energies ($\Delta E_{*H}$) for various combinations of 6 parameters on 9 HER-relevant metal surfaces. Our results are summarized in Table~\ref{tab:her_experiments} where each row is a different set of parameters and where columns correspond to a unique material and miller orientation. 

In order to verify that our calculations were correct, we first reproduced calculations using structures from the OC20 dataset and the exact set of parameters used in this dataset. We obtained $\Delta E_{*H}$ within 0.01~eV of what was reported (see the SI for more details). Therefore, it is interesting to note that when trying to reproduce other results from \citet{norskov_trends_2005} in Table~\ref{tab:her_experiments} where the exact structures were not available, we obtain differences of as much as 0.09~eV for Cu even though we used the same set parameters within what was reported.

Changing the exchange-correlation functional from the revised Perdew–Burke–Ernzerhof (RPBE) functional\cite{hammer1999improved} to the Perdew–Burke–Ernzerhof functional\cite{perdew1996generalized} makes the most difference with about $\sim$0.2~eV across all materials. Even though this difference is systematic, since, as we will discuss, the overpotential is related to the absolute value of the adsorption energy, it can change which material is predicted to be better. For example, the material with the smallest absolute value of $\Delta E_{*H}$ is Ag for PBE but it is Cu for RPBE when keeping all other parameters the same. Since RPBE has been shown to perform better on surface properties than PBE\cite{teng2014choosing, vega2018jacob} as it was designed to do, one could argue that RPBE should always be chosen. For that reason, we will not take this effect into consideration when quantifying uncertainty, but it is interesting to note given that many databases including OC22 use PBE.   

The coverage, how much of the surface is covered by adsorbates, has the second-largest effect on adsorption energy after the functional. Coverage can be controlled in two ways: by increasing the number of adsorbates on the surface (or slab) or by changing the size of the surface. The true coverage is not known in general and it may change as the reaction occurs. In reality all reaction intermediates exist simultaneously on the surface which makes this effect very difficult to model accurately with DFT.

The size of the reciprocal grid (number of k-points) and the plane wave energy cutoff should technically be chosen such that the energy is converged. However, in practice, convergence is relatively subjective and converging these parameters for every calculation is not feasible. Here we compared default parameters used in the OC20 dataset to slightly more expensive parameters and found that the energy cutoff had little effect, but that the k-point grid made a difference of about 0.06~eV alone on Pt. Moreover, it is common, as is done in OC20, to relax only certain layers of the slab to reduce computation when calculating energy. We found that relaxing more layers had almost no effect on adsorption energy.

Finally, adsorption sites, i.e. the position where adsorbates bind to the surface, play a key role in reaction kinetics \cite{norskov_fundamental_2014}. In computational studies, sites are typically chosen heuristically with respect to features of the surface (e.g. ``top'', ``bridge'', ``hollow'') with more recent studies sampling multiple possible sites to identify the ones with the lowest adsorption energy \cite{lan_adsorbml_2023}. However, the experimentally relevant sites are difficult or impossible to identify and may shift dynamically as the reaction proceeds. Here, we tested four different heuristic sites on Pt$\{$111$\}$ which is a simple flat surface and obtained a maximum difference of 0.05~eV. 

Overall, looking at the bottom of Table~\ref{tab:her_experiments} the maximum variation due to changing parameters was 0.3~eV when taking into account the functional and 0.17 eV when ignoring this effect. We did not test all possible combinations of parameters for all materials so it is very likely that the compound effect of these parameter choices is actually larger in practice.

\begin{table*}
\centering
\caption{\label{tab:her_experiments}Summary of all hydrogen adsorption energy calculations for uncertainty quantification. Default parameters are bolded. The kpoint grids are $a \times a \times 1$ where $a$ has the value in the table.}
\begin{adjustbox}{width=\linewidth}
\setlength\extrarowheight{2pt}
\begin{tabular}{lllllllgdgdgdgdg}
\toprule
\multicolumn{7}{c}{Parameters} & \multicolumn{9}{c}{Hydrogen adsorption energy (eV)} \\
Source & Func. & Cov. & Kpts & Encut & Relax & Sites & Pt\{111\} & Nb\{110\} & Ni\{111\} & Ir\{111\} & Cu\{111\} & Mo\{110\} & Co\{111\} & Pd\{111\} & Ag\{111\} \\
\midrule
\multirow[t]{2}{*}{\citet{norskov_trends_2005}} & \multirow[t]{2}{*}{RPBE} & 1/4 &  &  &  &  & -0.33 & -0.80 & -0.51 & -0.21 & -0.05 & -0.61 & -0.51 & -0.38 & 0.27 \\
\hhline{~~*{14}{-}}
 &  & 1 &  &  &  &  & -0.27 & -0.80 & -0.47 & -0.16 & 0.03 &  &  &  &  \\
\hhline{*{16}{-}}
\multirow[t]{13}{*}{Ours} & \multirow[t]{5}{*}{PBE} & \multirow[t]{2}{*}{1/4} & \multirow[t]{2}{*}{} & \textbf{350} &  &  & -0.47 & -0.90 & -0.69 & -0.44 & -0.3 & -0.78 & -0.69 & -0.55 & 0.11 \\
\hhline{~~~~*{12}{-}}
 &  &  &  & 450 &  &  & -0.48 &  &  &  &  &  &  &  &  \\
\hhline{~~*{14}{-}}
 &  & \multirow[t]{3}{*}{1} & \multirow[t]{2}{*}{\textbf{4}} & \textbf{350} &  &  & -0.40 & -0.85 & -0.65 & -0.37 & -0.18 &  &  &  &  \\
\hhline{~~~~*{12}{-}}
 &  &  &  & 450 &  &  & -0.39 &  &  &  &  &  &  &  &  \\
\hhline{~~~*{13}{-}}
 &  &  & 2 &  &  &  & -0.46 &  &  &  &  &  &  &  &  \\
\hhline{~*{15}{-}}
 & \multirow[t]{8}{*}{RPBE} & \multirow[t]{3}{*}{1/9} & 3 &  &  &  & -0.32 &  &  &  &  &  &  &  &  \\
\hhline{~~~*{13}{-}}
 &  &  & \multirow[t]{2}{*}{\textbf{4}} & \multirow[t]{2}{*}{} & \textbf{1} &  & -0.30 &  &  &  &  &  &  &  &  \\
\hhline{~~~~~*{11}{-}}
 &  &  &  &  & 3 &  & -0.28 &  &  &  &  &  &  &  &  \\
\hhline{~~*{14}{-}}
 &  & \multirow[t]{4}{*}{1/4} & \multirow[t]{4}{*}{} & \multirow[t]{4}{*}{} & \multirow[t]{4}{*}{} & \textbf{Oct-holes} & -0.23 & -0.75 & -0.47 & -0.28 & -0.14 & -0.65 & -0.53 & -0.37 & 0.27 \\
 &  &  &  &  &  & On-top & -0.25 &  &  &  &  &  &  &  &  \\
 &  &  &  &  &  & Bridge & -0.23 &  &  &  &  &  &  &  &  \\
 &  &  &  &  &  & Tet-holes & -0.20 &  &  &  &  &  &  &  &  \\
\hhline{~~*{14}{-}}
 &  & 1 &  &  &  &  & -0.20 & -0.73 & -0.42 & -0.21 & 0.03 &  &  &  &  \\
\midrule
 \multicolumn{7}{r}{Maximum difference} & 0.28 & 0.17 & 0.27 & 0.28 & \textbf{0.33} & 0.17 & 0.18 & 0.18 & 0.16 \\
 \multicolumn{7}{r}{Maximum difference (RPBE)} & 0.13 & 0.07 & 0.09 & 0.12 & \textbf{0.17} & 0.04 & 0.02 & 0.01 & 0 \\
\bottomrule
\end{tabular}
\end{adjustbox}
\end{table*}

\begin{table*}
\centering
\caption{\label{tab:oer_experiments}Summary of all oxygen evolution reaction adsorption energy calculations for uncertainty quantification. Default parameters are bolded.}
\begin{adjustbox}{width=\linewidth}
\setlength\extrarowheight{2pt}
\begin{tabular}{lllll*{4}{@{\hspace{8pt}}gdg}}
\toprule
\multicolumn{5}{c}{Parameters} & \multicolumn{12}{c}{Adsorption energy (eV)} \\
 &  &  &  &  & \multicolumn{3}{c}{SrTiO$_3$ \{100\}} & \multicolumn{3}{c}{SrCoO$_3$ \{100\}} & \multicolumn{3}{c}{IrO2 \{110\}} & \multicolumn{3}{c}{RuO$_2$ \{110\}} \\
Source & Cov. & Func. & Term. & Spin & O & OH & OOH & O & OH & OOH & O & OH & OOH & O & OH & OOH \\
\midrule
\citet{man_universality_2011} &  &  & & & 3.88 & 1.22 & 4.51 & 2.97 & 1.15 & 4.06 & 1.486 & -0.205 & 2.953 & 2.68 & 0.96 & 3.92 \\
\hhline{*{17}{-}}
\citet{gunasooriya_analysis_2020} &  &  &  &  &  &  &  &  &  &  & 1.61 & -0.12 & 3.02 &  &  &  \\
\hhline{*{17}{-}}
\multirow[t]{6}{*}{Ours} & \multirow[t]{4}{*}{\textbf{Default}} & \multirow[t]{4}{*}{} & \multirow[t]{2}{*}{Alt 1} & Off &  &  &  &  &  &  &  &  &  &  & 2.07 &  \\
 &  &  &  & \textbf{On} &  &  &  & 2.42 & 0.89 & 3.39 & 4.99 & 2.78 &  & 3.04 & 2.01 & 4.56 \\
\hhline{~~~*{14}{-}}
 &  &  & Alt 2 & &  &  &  &  &  &  & 1.06 & -0.21 &  &  &  &  \\
\hhline{~~~*{14}{-}}
&  & \textbf{RPBE} & \textbf{Default} & & 4.02 & 1.59 & 4.02 &  &  &  &  &  &  &  &  &  \\
\hhline{~~*{15}{-}}
 & & PBE & & & 4.10 & 1.48 &  &  &  &  &  &  &  &  &  &  \\
\hhline{~*{16}{-}}
 & Low  &  &  & & 3.62 & 1.12 &  & 2.73 & 1.39 & 3.87 & 1.49 & -0.12 & 2.87 &  &  &  \\
 \midrule
  \multicolumn{5}{r@{\hspace{8pt}}}{Maximum difference} & 0.48 & 0.47 & 0.49 & 0.55 & 0.50 & 0.67 & \textbf{3.93} & 2.99 & 0.15 & 0.36 & 1.11 & 0.64 \\
 \multicolumn{5}{r@{\hspace{8pt}}}{Maximum difference (Default term.)} & 0.48 & 0.47 & \textbf{0.49} & 0.24 & 0.24 & 0.19 & 0.12 & 0.08 & 0.15 & & &  \\
\bottomrule
\end{tabular}
\end{adjustbox}
\end{table*}

\subsection{OER calculations} \label{sec:oer_calcs}

Catalysts for HER are typically transition metal surfaces which is what can be found in the OC20 dataset. However, for the OER, the experimental conditions are highly oxidizing and often acidic, so more complex oxides are typically used because they remain stable under those conditions. Computing $\eta$ for OER requires three adsorption energies: $\Delta G_{*OH}$, $\Delta G_{*O}$ and $\Delta G_{*OOH}$. We calculated these three adsorption energies for combinations of 4 parameters on 4 different oxide surfaces. Our results are summarized in Table~\ref{tab:oer_experiments} where each row is a different set of parameters and where columns correspond to a unique adsorbate, material and miller orientation.

Coverage is more difficult to define on more complex surfaces so it is hard to obtain the exact same coverage as literature data. Therefore, the difference in Table~\ref{tab:oer_experiments} between our values and literature (0.1-0.2~eV) and the difference between different literature values (0.1~eV) are probably due mostly to a difference in coverage and site choices. When manually testing the effect of coverage on SrTiO$_3$, we obtained large differences of as much as 0.47~eV for *OH between low and high coverages.   

For the OER, we only tested the effect of the functional on the SrTiO$_3$$\{$100$\}$ surface. We found that it was slightly smaller than for HER surfaces (0.1~eV) and not as systematic: PBE leads to a higher energy for *O compared to RPBE whereas it is the opposite for *OH. The choice of functional is more important here since OC22 uses PBE and this is the data we will test the metric on. 

Surface termination refers to the way the surface is cut and prepared before the adsorbate is placed on it. Depending on the material and its symmetry, for a specific set of Miller indices, there are multiple ways to prepare the surface. The termination, i.e., which atoms end up at the surface, can lead to important surface reorganization and to the creation of surface states that completely change calculated adsorption energies especially in oxides. Here, we estimated the effect of surface termination on IrO$_2$ by computing adsorption energies for three different terminations: a symmetric corrugated surface with minimal undercoordination (expected), an O-terminated surface and an Ir-terminated surface. Unsurprisingly, this caused differences of as much as $\sim$3~eV with some *OOH calculations not able to converge. Of course, it is possible to carefully choose terminations to avoid these discrepancies, but finding realistic terminations can require in-depth analysis \cite{therrien2021theoretical, stevanovic2014variations} that is impractical for high throughput studies. This may explain why terminations were chosen randomly in OC22. 

Since incorrect terminations lead to extreme differences in adsorption energies that could, in principle, be discarded automatically and that it is unlikely that an unstable surface lead to a good $\eta$, we will not directly consider the very large effect ($\sim$4~eV) of termination as measured in Table~\ref{tab:oer_experiments} in our uncertainty estimate.  Overall, looking at the bottom of Table~\ref{tab:oer_experiments} the maximum variation due to changing parameters was 3.93~eV when taking into account the termination and 0.49 eV when ignoring this effect. The differences were particularly large on SrTiO$_3$ because of the high coverage experiment. Again, we did not test all possible combinations of parameters for all materials so it is very likely that the compound effect of these parameter choices is actually larger in practice. Moreover, although we ignore the extreme effect of unstable terminations, actual valid alternate terminations for a surface may exist and cause differences that are not measured here.

\subsection{Statistical analysis of OC20 and OC22 repeat surfaces} \label{sec:stats}

The effect of terminations, coverage and sites which cause some of the largest differences in energy can be measured statistically by analyzing entries in OC20 and OC22 that have the same bulk structure and the same miller indices. The results of that analysis for the HER adsorbate and the three OER adsorbates are presented in Table~\ref{tab:similar}.

\begin{table}
\begin{minipage}{0.48\textwidth}
\centering
\caption{Statistical analysis of structures with the same miller indices in OC20 and OC22. RMSE$^*$ is the room mean square difference with the average calculated adsorption energy of each group in eV.}
\label{tab:similar}
\begin{adjustbox}{width=\linewidth}
\begin{tabular}{llg@{\hspace{8pt}}dgdg}
\toprule
 &  & \multicolumn{1}{c@{\hspace{8pt}}}{OC20} & \multicolumn{4}{c}{OC22} \\
 &  & *H & *OH & *O & *OOH & All \\
\midrule
\multirow[t]{3}{0.3\linewidth}{Same miller (termination + coverage + site)} & $n_{\text{groups}}$ / $n_{\text{struc.}}$ & 55 / 109 & 276 / 629 & 1808 / 4683 & 114 / 262 & 2198 / 5574 \\
 & RMSE$^*$ (eV) & 0.29 & 1.02 & 0.97 & 0.69 & 0.96 \\
 & Max diff (eV) & 1.57 & 9.84 & 13.84 & 4.82 & 13.84 \\
\midrule
\multirow[t]{3}{0.3\linewidth}{Same miller and $n_\text{ads}$ (termination + site)} & $n_{\text{groups}}$ / $n_{\text{struc.}}$ & & 191 / 428 & 1036 / 2468 & 113 / 260 & 1340 / 3156 \\
 & RMSE$^*$ (eV) & & 1.10 & 0.87 & 0.69 & 0.89 \\
 & Max diff (eV) & & 9.84 & 11.58 & 4.82 & 11.58 \\
\midrule
\multirow[t]{3}{0.3\linewidth}{Same slab ID (coverage + site)} & $n_{\text{groups}}$ / $n_{\text{struc.}}$ & & 22 / 44 & 1363 / 2819 & 13 / 26 & 1398 / 2889 \\
 & RMSE$^*$ (eV) & & 0.76 & 0.77 & 0.42 & 0.76 \\
 & Max diff (eV) & & 4.66 & 11.58 & 1.86 & 11.58 \\
\midrule
\multirow[t]{3}{0.3\linewidth}{Same slab ID and $n_\text{ads}$ (site)} & $n_{\text{groups}}$ / $n_{\text{struc.}}$ & & 12 / 24 & 518 / 1051 & 13 / 26 & 543 / 1101 \\
 & RMSE$^*$ (eV) & & 0.58 & 0.71 & 0.42 & 0.71 \\
 & Max diff (eV) & & 2.86 & 11.58 & 1.86 & 11.58 \\
\bottomrule
\end{tabular}
\end{adjustbox}
\end{minipage}
\end{table}

There are only 55 surfaces with a hydrogen adsorbate in OC20 that have repeat calculations with on average about 2 calculations per surface. The difference between these two calculations could be due to the termination or the site, not the coverage since there is always only one adsorbate per surface in OC20. The average root mean squared distance between values and the mean of each set (RMSE$^*$) is 0.29 eV which means that if a structure has two calculations they would each be, on average, 0.29~eV from the mean, i.e. 0.6~eV apart.

OC22 has a lot more repeat surfaces that may or may not have the same coverage or termination. When looking at surfaces that only share miller indices, the RMSE$^*$ is nearly 1~eV on average across all three adsorbates. When considering only calculations with the exact same parameters except the adsorption site, RMSEs go down to about 0.7~eV. These values are very large considering that, as we discuss before there is no consensus on what constitute the true site coverage or termination and that it can be materials specific. Even if one were to gather all these calculation for each surface, there is no agreed upon way to combine them into a single predictive number.

\subsection{Gibbs free energy correction} \label{sec:gibbs_corr}

To predict the overpotential, one actually needs the adsorption \textit{free} energy at operating conditions, because what dictates the reaction kinetics at constant pressure and temperature is the Gibbs free energy. It is very common to use a constant Gibbs free energy correction to account for this effect. The correction calculated by \citet{norskov_trends_2005} is used throughout the literature as the default free energy correction for *H even though it was computed with a specific set of parameters on a single copper surface. We computed Gibbs free energy correction for several surfaces with different parameters (see Table~\ref{tab:gibbs_corr} and found that the Gibbs free energy correction alone could vary by as much as 0.1~eV for *H, 0.2~eV for *O and *OH and 0.3~eV for *OOH.

\begin{table}
\begin{minipage}{0.48\textwidth} 
    \centering
    \caption{Gibbs Free energy corrections ($\Delta G_{corr}$) in eV for different adsorbates on various surfaces. Default parameters are bolded.}
    \label{tab:gibbs_corr}
    \begin{adjustbox}{width=\linewidth}
\setlength\extrarowheight{2pt}
\begin{tabular}{llllgdgd}
\toprule
Source & Material & Func. & Cov. & H & OH & O & OOH \\
\midrule
\citet{tran_open_2023}(OC22) & Average & PBE &  &  & 0.26 & -0.03 & 0.22 \\
\hline
\multirow[t]{2}{0.3\linewidth}{\citet{gunasooriya_analysis_2020}} & Average &  &  &  & 0.295 & 0.044 & 0.377 \\
& & & & & & & \\
\hline
\citet{norskov_trends_2005} & Cu \{111\} &  &  & 0.24 &  &  &  \\
\hline
\multirow[t]{9}{*}{Ours} & \multirow[t]{2}{*}{SrTiO$_3$ \{100\}} & \textbf{RPBE} &  &  & 0.103 & -0.189 & 0.036 \\
\hhline{~~*{6}{-}}
&  & PBE &  &  & 0.119 & -0.195 &  \\
\hhline{~*{7}{-}}
 & SrCoO$_3$ \{100\} & &  &  & 0.296 & -0.088 & 0.131 \\
\hhline{~*{7}{-}}
 & IrO$_2$ \{110\} & &  &  & 0.221 & -0.079 & 0.164 \\
\hhline{~*{7}{-}}
 & \multirow[t]{2}{*}{Pt \{111\}} & \multirow[t]{2}{*}{} & 1/4 & 0.13 &  &  &  \\
 &  &  & 1 & 0.15 &  &  &  \\
\hhline{~*{7}{-}}
 & \multirow[t]{2}{*}{Cu \{111\}} & \multirow[t]{2}{*}{} & 1/4 & 0.14 &  &  &  \\
 &  &  & 1 & 0.2 &  &  &  \\
\hhline{~*{7}{-}}
 & Mo \{111\} &  & 1/4 & 0.19 &  &  &  \\
\midrule
\multicolumn{4}{r}{Maximum difference} & 0.11 & 0.19 & 0.24 & \textbf{0.34} \\ 
\bottomrule
\end{tabular}
\end{adjustbox}
\end{minipage}
\end{table}

\end{document}